\documentclass[prx,aps,amssymb,shownopacs,twocolumn,10pt,floatfix]{revtex4-1}
\usepackage{amsmath}
\usepackage{amssymb}
\usepackage{amsthm}
\usepackage{amsfonts}
\usepackage{float}
\usepackage{listings}
\usepackage{enumerate}
\usepackage{latexsym}
\usepackage{hyperref}
\usepackage{braket}
\usepackage{natbib}
\usepackage{psfrag}
\usepackage{color}
\usepackage{bm}
\usepackage[pdftex]{graphicx}
\usepackage{subfigure}

\def\matrix22#1#2#3#4{\left(\begin{array}{cc}#1&#2\\#3&#4\end{array}\right)}

\usepackage{ulem}

\begin{document}

\title{Characterizing the many-body localization transition using the entanglement spectrum}

\author{S. D. Geraedts$^1$}
\author{N. Regnault$^2$}
\author{R. M. Nandkishore$^3$}
\affiliation{$^1$ Department of Electrical Engineering, Princeton University, Princeton NJ 08544, USA\\
$^2$ Laboratoire Pierre Aigrain, Ecole Normale Sup\'erieure-PSL Research
University, CNRS, Universit\'e Pierre et Marie Curie-Sorbonne Universit\'es,
Universit\'e Paris Diderot-Sorbonne Paris Cit\'e, 24 rue Lhomond, 75231
Paris Cedex 05, France\\
$^3$ Department of Physics and Center for Theory of Quantum Matter, University of Colorado at Boulder, Boulder CO 80309, USA}

\begin{abstract}
We numerically explore the many body localization (MBL) transition through the lens of the {\it entanglement spectrum}. While a direct transition from localization to thermalization is believed to obtain in the thermodynamic limit (the exact details of which remain an open problem), in finite system sizes there exists an intermediate `quantum critical' regime. Previous numerical investigations have explored the crossover from thermalization to criticality, and have used this to place a numerical {\it lower} bound on the critical disorder strength for MBL. A careful analysis of the {\it high energy} part of the entanglement spectrum (which contains universal information about the critical point) allows us to make the first ever observation in exact numerics of the crossover from criticality to MBL and hence to place a numerical {\it upper bound} on the critical disorder strength for MBL. 
\end{abstract}

\date{\today}
\maketitle

\paragraph*{Introduction}
Many body localization (MBL) and the resulting breakdown of statistical mechanics in disordered interacting systems has been the subject of much recent research \cite{Gornyi, BAA,Znidaric, OganesyanHuse,Pal,Nandkishore-2015}. The intensive research has yielded a plethora of insights into the properties of this non-ergodic regime, including its connections of integrability \cite{Bardarson2012, Serbyn, HNO, Scardicchio, lstarbits, GBN}, its response properties \cite{nonlocal, mblconductivity}, and the circumstances under which the phenomenon may arise \cite{QHMBL, 2dcontinuum, anycontinuum, proximity, nonabelian, mblmobilityedges, mblbathgeneral, avalanches}. However, the quantum phase transition between a `thermal' phase where statistical mechanics is obeyed and a `many body localized' (MBL) phase where it is not continues to be an open problem. This is a {\it dynamical} transition which lies outside the usual thermodynamic frameworks, and while it has been attacked with a variety of techniques, from mean field theory \cite{lauman, gn} to the strong disorder renormalization group \cite{vho, pvp, huse2, potter2}, and using both analytic arguments \cite{grover, SerbynMoore, clo} and numerics \cite{SerbynThouless, Luitz2015, Khemani2016}, (see \cite{pvp3} for a review), a complete understanding of the transition remains elusive. 

\begin{figure}[tbh]
\includegraphics[width=\linewidth]{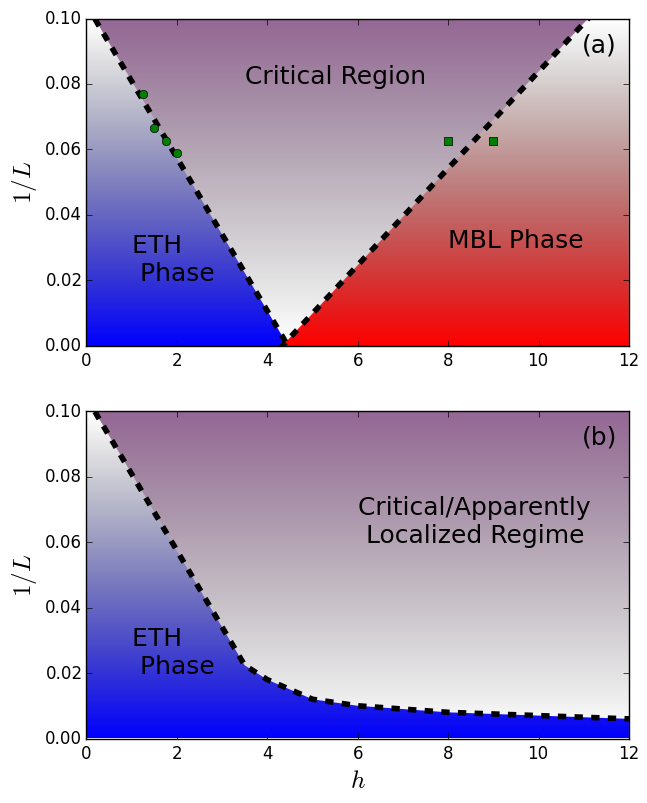}
\caption{Possible phase diagrams for the thermal-MBL transition, as a function of disorder strength $h$ and system size $L$. 
The phase diagram proposed by Ref. \onlinecite{Khemani2016} is given in (a), in the thermodynamic limit it has a direct transition from the thermal to MBL phases, but at the finite sizes accessible numerically there is an intermediate `critical' regime.
Fig. (b) shows an alternative phase diagram in which there is no MBL phase in the thermodynamic limit, though finite sized systems may be glassy. 
Previous numerical investigations have only observed transition between the thermal and critical phases, and therefore cannot determine which of these phase diagrams are correct.
Our examination of the eigenstate entanglement spectrum allows us to see {\it both} sides of the quantum critical fan, thus placing both lower and  {\it upper} bounds on the critical disorder strength, and establishing (a) as the correct phase diagram.
The symbols correspond to where we have detected a transition out of the critical fan using the entanglement spectrum.
}
\label{phase}
\end{figure}
Many investigations have focused on the (Von Neuman) {\it eigenstate entanglement entropy} (EEE) as an order parameter for the phase transition. An early analytic paper argued \cite{grover} that if the EEE density evolved in a smooth fashion across the MBL transition, then the critical point must exhibit thermal entanglement. However, a careful parsing of results from numerical exact diagonalization \cite{Khemani2016} suggests a different picture, whereby the EEE density is {\it discontinuous} at the transition, in the thermodynamic limit. This work \cite{Khemani2016} put forth an appealing picture wherein the MBL and thermal phases are separated in finite size numerics by a `quantum critical fan' [See Fig.~\ref{phase}(a)], just like thermodynamic quantum critical points \cite{Sachdev}. However, exact numerical investigations to date have only seen the left side of the fan i.e. the crossover from the thermal phase to the quantum critical regime. As a result, not only is the global structure of the phase diagram still unresolved, but since the disorder strength $h_t$ for the thermal to critical crossover increases with system size, existing exact numerics are only able to place a {\it lower bound} on the critical disorder strength for the MBL transition \footnote{The transition also occurs as a function of energy density, in this work we will fix the energy density to be in the middle of the density of states.}. This lower bound has drifted gradually upwards as numerics on larger system sizes have become available, and while finite size scaling collapses (e.g. \cite{Luitz2015}) suggest a finite critical disorder strength,  it is possible that the sizes accessed are not large enough to be in the scaling regime, a perspective reinforced by the observation that the critical exponents extracted from such analysis violate analytic bounds \cite{clo} and are therefore likely incorrect. Existing exact numerics thus cannot exclude the possibility that $h_t\rightarrow\infty$ as $L\rightarrow\infty$, i.e. that the MBL phase observed in numerics is nothing more than a finite-size effect, and the correct phase diagram resembles Fig.~\ref{phase}(b). A direct observation of the MBL to critical crossover using exact numerical methods is thus highly desirable, both to confirm the global structure of the phase diagram, and to enable the extraction of a numerical {\it upper} bound on the critical disorder strength for the MBL transition.

In this work we analyze the thermal-MBL transition through a new lens, that of the eigenstate entanglement spectrum (EES) \cite{LiHaldane, Chamon, Geraedts2016, SAP, Chamon2, Pietracaprina}. The EES contains far more information about the pattern of quantum entanglement than the EEE. In earlier work \cite{Geraedts2016}, we revealed a rich and universal structure in the EES of the MBL phase, and pointed out that the {\it high energy} part of the EES (the part typically discarded in numerical analyses) contains robust information about the critical regime. We now parse the system size dependence of the high energy EES - and find this allows us to detect {\it both} the thermal-critical and MBL-critical crossovers. Our work thus not only confirms the scenario of a `quantum critical fan' in finite size numerics, but also allows us to {\it upper bound} the critical disorder strength for the MBL transition. 

\paragraph*{Model}
In this work we perform exact diagonalization on a spin one half Heisenberg model of length $L$ with random fields in the $x$ and $z$ directions and periodic boundary conditions:
\begin{equation}
H=J\sum_{i=1}^L \vec S_i \cdot \vec S_{i+1} + h\sum_i (\alpha_{iz} S_i^z + \alpha_{ix} S_i^x),
\label{H}
\end{equation}
where $2S_\mu= \sigma^\mu$ are the Pauli matrices.
Here we set $J=1$ while $h$ controls the disorder strength. The $\alpha_{i\mu}$ are random numbers drawn from a uniform distribution in $[-1,1]$. This model is known to have a thermal-critical transition around $h=2.5$ \cite{Geraedts2016}. Eq.~(\ref{H}) is similar to the `standard model' used in studies of many-body localization which has a random field in only one direction. We use two random fields to break the total $S^z$ conservation. 
Without breaking this symmetry, we have to choose a total $S^z$ sector for every entanglement spectra. Different sectors are available depending on whether $L$
\footnote{and also $L_A$}
is even or odd, and this leads to significant even/odd effects in our data. Breaking the symmetry by introducing a field along the x axis eliminates these effects, allowing us to compare even and odd $L$ and therefore doubling our resolution in system size. 

We use exact diagonalization to obtain the eigenstates of Eq.~(\ref{H}) for system sizes in the range $L=11-17$, for $200-500$ realizations of disorder. Extensive disorder averaging is necessary, since sample to sample fluctuations are known to be large close to the transition \cite{Khemani2016}. For $L\le 15$ we compute all the eigenstates of the system. However since the entanglement properties depend on the energy density we use only the middle $1/3$ of these eigenstates in our calculations, similar to \cite{Pal, Geraedts2016}. For $L>15$ we use the shift-invert method to obtain $1000$ eigenstates at energy densities approximately halfway between the top and bottom of the spectrum.

\paragraph*{Entanglement Density of States}
Having obtained the eigenstates the next task is to extract from them some useful information. Entanglement has long been understood to be a useful quantity to probe MBL.  Historically work focuses on the entanglement entropy, defined as:
\begin{equation}
S=- {\rm Tr}(\rho_A\log\rho_A); ~~~ \rho_A={\rm Tr}_B|\Psi\rangle\langle\Psi|,
\end{equation}
where $|\Psi\rangle\langle\Psi|$ is the density matrix of an eigenstate, and $\rho_A$ is the `reduced density matrix' for subregion $A$ obtained by tracing out all the degrees of freedom in some region $B$ which is the complement to $A$
\footnote{What we have defined is the Von Neumann entanglement entropy, other entanglement entropies (e.g. Renyi) are defined similarly.}.
The subregion $A$ is composed of $L_A$ consecutive sites.
Entanglement entropy exhibits area law behavior in the MBL phase and volume law behavior in the thermal phase, making it a useful probe of the physics of the transition. 
A potential drawback of it is that it distills a lot of information (e.g. the entire matrix $\rho_A$) down to a single number. More information can be potentially found in the entanglement spectrum, $\{\lambda_i \}$, which is the set of all eigenvalues of $-\log\rho_A$. 
The entanglement entropy is dominated by `entanglement states' with low `entanglement energy' (i.e.~small $\lambda_i$). However, 
in a previous work \cite{Geraedts2016} we showed that entanglement states  at high entanglement energy have universal structure in the MBL phase, and appear to carry a memory of the critical point. Motivated by this observation, we focus in the present work on the {\it high energy} part of the entanglement spectrum to tease out insights into the MBL transition.  Such information can only be obtained by studying the spectrum directly since these states' contribution to the entanglement entropy is small. 
\begin{figure}[tbh]
\includegraphics[width=\linewidth]{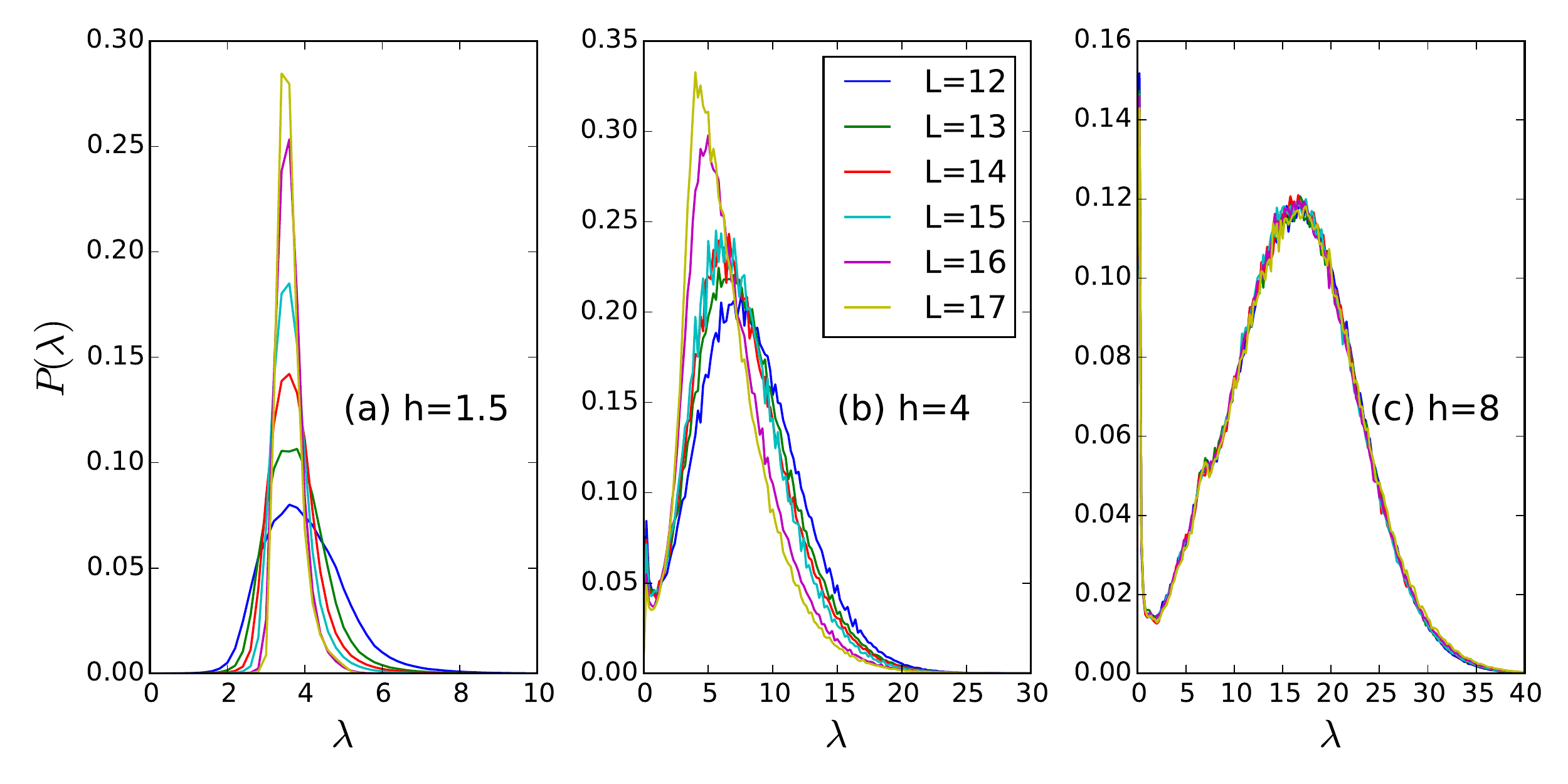}
\caption{Entanglement density of states (EDOS) for phases in the thermal phase with $h=1.5$ (a), in the critical phase with $h=4$ (b), and in the MBL phase with $h=8$ (c), for $L_A=5$. In the thermal phase the EDOS is peaked around a value proportional to $L_A$ and independent of $L$. In the MBL phase the EDOS has two peaks, one which exponentially decays from zero and the other at higher entanglement energies. In the critical phase the system moves between the two cases as a function of $L$. The EDOs shown we obtained by averaging over the middle third of the  eigenstates in $128-512$ realizations of disorder.}
\label{dos2}
\end{figure}

\begin{figure}[tbh]
\includegraphics[width=\linewidth]{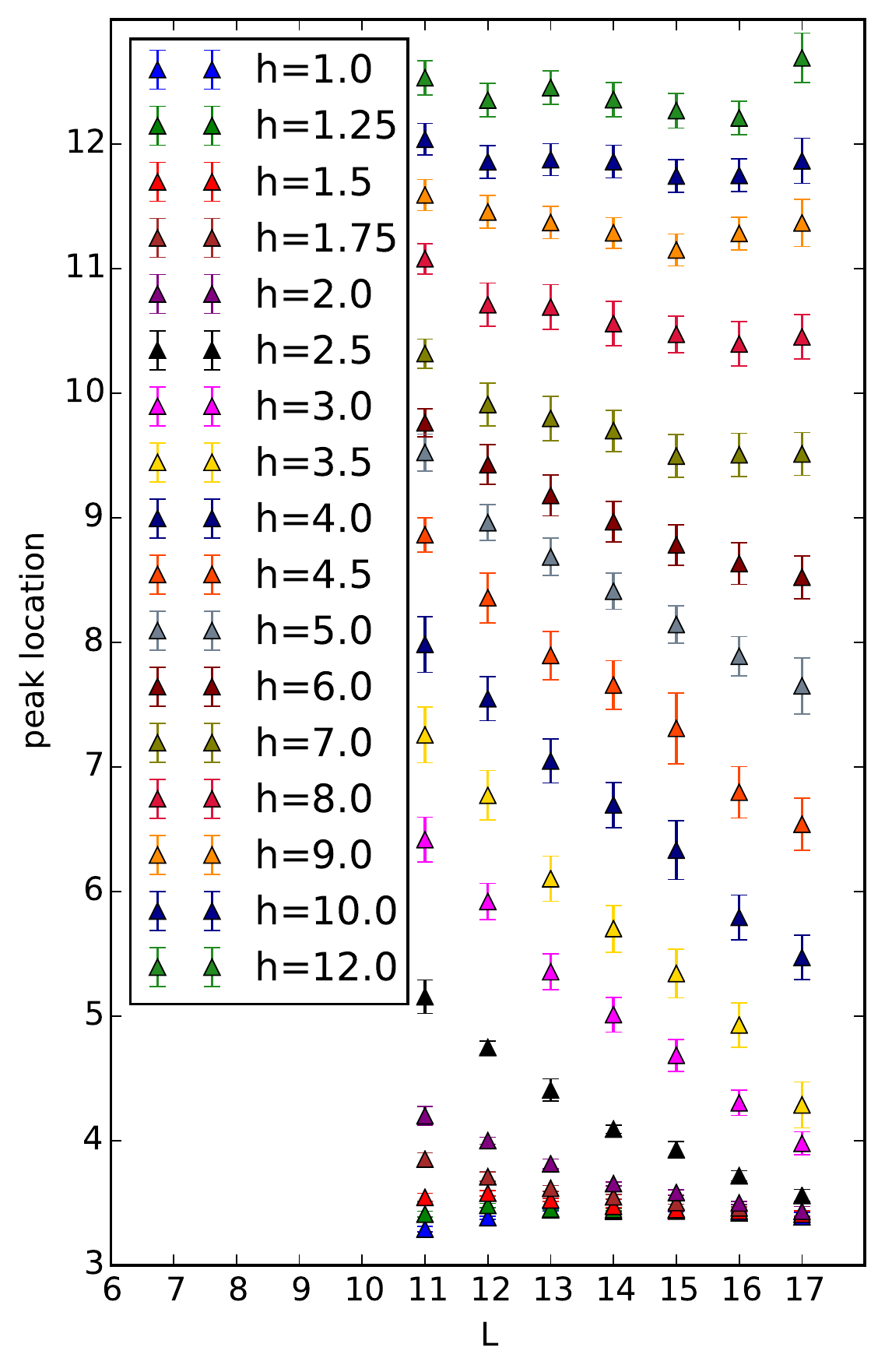}
\caption{The location of the maximum of the EDOS (excluding the peak around $\lambda=0$ which occurs in the MBL phase), as a function of $L$ for several $h$ values and $L_A=5$. The peak location should only depend on $L$ in the critical regime. Thus this plot allows us to identify both the critical-thermal boundary and the critical-MBL boundary.  The peak location for a single sample was computed by averaging maximum values of the EDOS of all that sample's eigenstates (see Appendix \ref{app::avg} for justification of this procedure). The plotted data and errors are the result of averaging the peak locations of $128-512$ samples. (More samples were used at smaller sizes and larger disorder.)
The separation between curves at different $h$ is larger than the change in the curves with $L$, especially at large disorder. Therefore to show all data clearly on the same plot we subtracted a constant from each curve with $h\ge 3.5$. The following is a list of the values of to constants which were subtracted in the format (h:offset): 
(3.5:0.5), (4:1), (4.5:1.5), (5:2), (6:4), (7:5.5), (8:6.5), (9:7.5), (10:8.5), (12:10.5)
}
\label{peaks}
\end{figure}

The quantity we will be interested in is the density of states (EDOS) of the entanglement spectrum. Fig.~\ref{dos2} shows examples of the EDOS in the ETH (a) and MBL (c) phases. To understand the ETH density of states, the prototypical example is a system at infinite temperature. In this case every eigenvalue of the reduced density matrix is equal. Let us define $N_A$ to be the size of the Hilbert space of the traced-out region $A$, and $L_A$ to be the length of this region in the one-dimentional systems we will study. 
Then $\lambda_i= \log N_A \propto L_A$ for all $i$. Since the systems we study are not at infinite temperature we instead get a density of states narrowly distributed around $L_A$. 
In the MBL phase a model wavefunction could be a pure product state which has $\lambda_0=0$ and all other $\lambda_i=\infty$. 
Away from this ideal limit we believe that the eigenstates should be products of `l-bits', which decay exponentially in space. Therefore there will be some non-zero $\lambda$ values which result from cutting the tails of these l-bits. From this intuition we expect an EDOS which decays exponentially from $\lambda=0$.
However in~\cite{Geraedts2016} we showed that the majority of entanglement eigenvalues are not in this exponentially decaying part but are instead in another peak which is at much higher entanglement energy. 
The level spacings in this higher peak obey semi-Poisson statistics, which are indicative of criticality. The location of this `high-energy peak' turns out to be a very sensitive probe of the critical properties of the system. We have studied the scaling of the `high energy peak' with disorder strength $h$, subsystem size $L_A$, and system size $L$. In the main text we study the dependence on system size $L$, which turns out to be the most informative, and allows us to detect the both the thermal-critical and critical-MBL transitions. The scaling with $h$ and $L_A$ is discussed in the Appendix. Note that we found that we need $L_A \ge 5$ in order to have good statistics. Since for any location of entanglement cut, $L_A$ is the {\it smaller} of the two subregions, it then follows that we need $L \ge 11$. Meanwhile, the largest system sizes we can access given available numerical resources are $L=17$, and thus our results are limited to the range $11 \le L \le 17$. A larger range may be accessible with specifically tailored models (e.g. \cite{husemodel}) or with greater computational resources, and would be an interesting problem for future work.

\paragraph*{Results}

Deep in either the thermal or MBL phases, we expect the EDOS to be independent of $L$. In the thermal phase, our intuition from the infinite-temperature case tells us that the EDOS should depend only on the number of degrees of freedom in the traced-out region $A$, i.e. it depends on $L_A$ but not on $L$. In the MBL phase the correlation length is very small. Increasing $L$ for fixed $L_A$ means adding degrees of freedom far from the region $A$, and these should not affect the entanglement between $A$ and the rest of the system. In Fig.~\ref{dos2} we show how the EDOS changes with system sizes for a few different disorder strengths. Deep in the thermal (a) and MBL phases (c) the EDOS peak position changes little with $L$, as expected. But at intermediate values of $h$ (b), which we might expect to lie in the critical region, we see significant changes in the EDOS peak position as a function of $L$.

The dependence of the EDOS in the critical region has a natural explanation in terms of the scenario outlined in \cite{Khemani2016} (and anticipated also in \cite{pvp}). In that work it was argued that entanglement in the critical regime is dominated by sparse `fractal' networks of long range resonances. Adding degrees of freedom far away from the entanglement cut alters the structures of these resonant networks, and hence alters the EDOS. As we increase system size at subcritical disorder, tuning through the critical to thermal crossover, these `resonant networks' densely fill in, so that EDOS becomes insensitive to system size as discussed above. Meanwhile, as we increase system size at supercritical disorder, tuning through the critical to MBL crossover, then eventually system size exceeds the size of the resonances, such that adding degrees of freedom far from the cut no longer effects the resonant networks across the entanglement cut, or the EDOS. The key diagnostic of criticality is thus a dependence of the EDOS on total system size $L$, at constant $h$ and $L_A$. 

We can quantify the motion of the density of states with system size by tracking the position of the top of the high entanglement energy peak. In Fig.~\ref{peaks} we plot this peak location as a function of $L$ for a variety of disorder strengths, at $L_A=5$. Note that the curves in this plot correspond to moving down vertical lines in Fig.~\ref{phase}. At small or large $h$ the curves are flat, since at the system sizes we can access these disorder strengths are entirely in either the thermal or MBL phases. At intermediate disorder, e.g. $h=3$, we see curves which decrease steadily, implying that at the system sizes we can access these curves are all in the critical regime. We can also see curves, e.g. at $h=2$ or $h=7$, which have a system size dependence only at small sizes. These curves indicate transitions between the critical regime and either the thermal or MBL phases. 

The crossover lengthscale $L_c(h)$ for the quantum critical fan is the lengthscale where the dependence of the EDOS peak on $L$ disappears. Note that we observe a crossover lengthscale at weak disorder, which is an increasing function of $h$ (corresponding to the left side of the quantum critical fan), and also a crossover lengthscale at strong disorder. While our resolution is limited by the small range of system sizes we can access numerically, it is clear that $L_c(h=6) > 17$, $L_c(h=7,8,9) = 15 \pm 1$, and $L_c(h=12) \lesssim 12$. The expert reader may wonder if a change in behavior around $L=15$ may be due to a change from full diagonalization to a shift-invert procedure, but we have verified that it is in fact physical, and not a numerical artifact (see Appendix \ref{app::avg}).   This second crossover lengthscale, which is a {\it decreasing} function of disorder strength, corresponds to the right side of the quantum critical fan in Fig.1, constituting to our knowledge the first observation of the critical to MBL crossover. Note also that insofar as $h=7$ lies clearly on the right side of the quantum critical fan, we can place a numerical {\it upper bound} on the critical disorder strength in this model, $h_c < 7$, to complement the lower bound $h_c \ge 2.5$ from the thermal to critical crossover. While the bounds $2.5 < h_c < 7$ are not particularly tight, the fact that we can place an upper bound at all is, in our view, a major advance - a further tightening of the bounds now requires `merely' the application of greater (or more cleverly applied) computing power. We also believe that the apparent `uptick' in the peak position at large $h$ and largest $L$ is a numerical artifact (in this regime the curves are `flat within error bars'), although it is interesting to speculate about a possible second crossover, which could modify the picture of the phase diagram in Fig.1. 

 {\it Conclusions}: We have examined the MBL transition through the lens of the entanglement spectrum. In particular, we have shown that {\it system size dependence} of the {\it high energy peak} in the entanglement spectrum is a sensitive measure of criticality, which allows us to identify both the critical to thermal crossover (which many studies have seen before), and the critical to MBL crossover (which has never before been observed to our knowledge). As such, not only does our work justify a picture in which at finite size the MBL and thermal regimes are separated by a `critical' fan, but it also allows us to place both a lower and an {\it upper} bound on the critical disorder strength for the MBL transition. A further tightening of these bounds, either by the application of greater computational resources or by the refinement of techniques introduced in this paper, or by some other method entirely, would be an interesting challenge for future work. Also interesting would be to apply similar analyses for models with quasi-periodic disorder, which may also be relevant for understanding the MBL transition \cite{Khemaniquasiperiodic}, or to examine the dynamical evolution of the entanglement spectrum starting from low entanglement initial conditions, as pioneered by \cite{Chamondynamics}. These problems too we leave to future work.

{\bf Acknowledgements} We thank S.A. Parameswaran, R. Vasseur, A.C. Potter, R. N. Bhatt and especially Vedika Khemani for feedback on the manuscript. R.N. is supported by the Foundational Questions Institute (fqxi.org; grant no. FQXi-RFP-1617) through their fund at the Silicon Valley Community Foundation. S.D.G. is supported by Department of Energy  BES Grant DE-SC0002140.

\begin{appendix}

\section{Dependence of entanglement spectra on subsystem size and disorder strength}
\label{app::depend}

In the main text we focussed on the dependence of the entanglement density of states on system size, since this is the quantity which reveals the systems critical properties. 
In this Appendix we check our understanding of the EDOS by confirming that it behaves correctly as a function of $L_A$, the subsystem size, and $h$, the disorder strength. 

\begin{figure}[tbh]
\includegraphics[width=\linewidth]{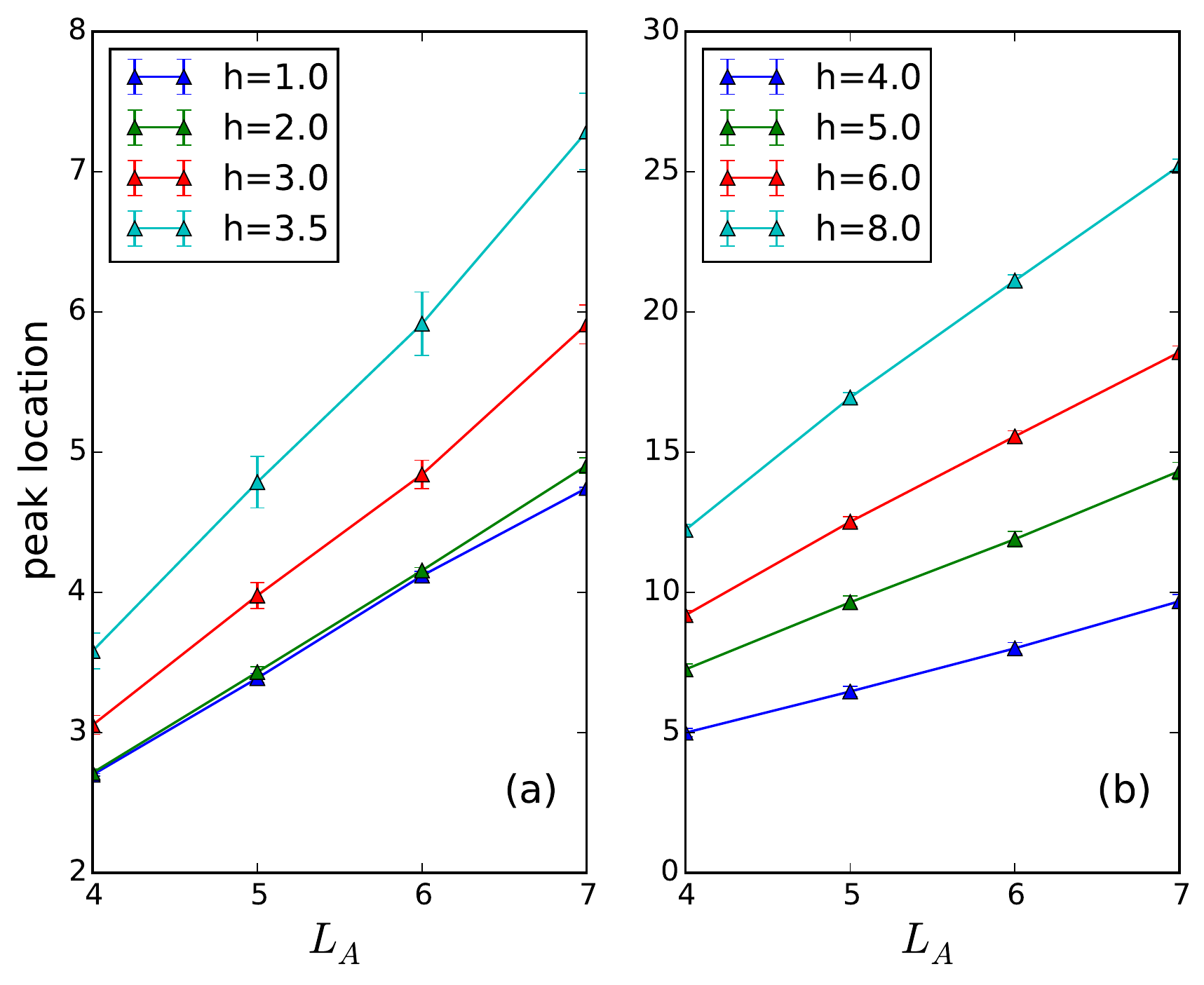}
\caption{The location of the maximum of the EDOS (excluding the peak around $\lambda=0$ which occurs in the MBL phase), as a function of $L_A$ for several $h$ values and $L=16$. As discussed in the main text, the peak location grows $\propto L_A$ in all phases. The data has been broken into multiple panels to make the small $h$ behavior easy to see.}
\label{fig::L_A}
\end{figure}

First we study the dependence on $L_A$. As discussed in the main text, in the thermal phase the location of the peak in the EDOS should be $\propto L_A$. In the MBL phase, we also expect the peak's location to scale as $L_A$, by the following argument: most degrees of freedom live a distance $\sim L_A$ away from the entanglement cut. (This is just dimensional analysis -  $L_A$ is the only lengthscale characterizing the cut). However, in the eigenstates degrees of freedom are typically only entangled within a lengthscale $\xi$, where $\xi$ is the localization length. Given exponential localization, a degree of freedom a distance $\sim L_A$ away from the cut will have entanglement $\sim \exp(-L_A/\xi)$ across the cut, and thus will have entanglement energy $\sim L_A/\xi$. 
In Fig.~\ref{fig::L_A} we show the peak location, as a function of $L_A$ for fixed $L=17$ and several disorder strengths. We see that in the thermal phase the behavior is $\propto L_A$ as expected [Fig.~\ref{fig::L_A}(a)]. In the MBL phase the peak location scales approximately as $L_A$, but there are small deviations from perfect linear behavior in the deep MBL regime[Fig.~\ref{fig::L_A}(b)]. Understanding whether these deviations are physical, and to what physics they correspond if so, is an interesting challenge for future work. 

\begin{figure}[tbh]
\includegraphics[width=\linewidth]{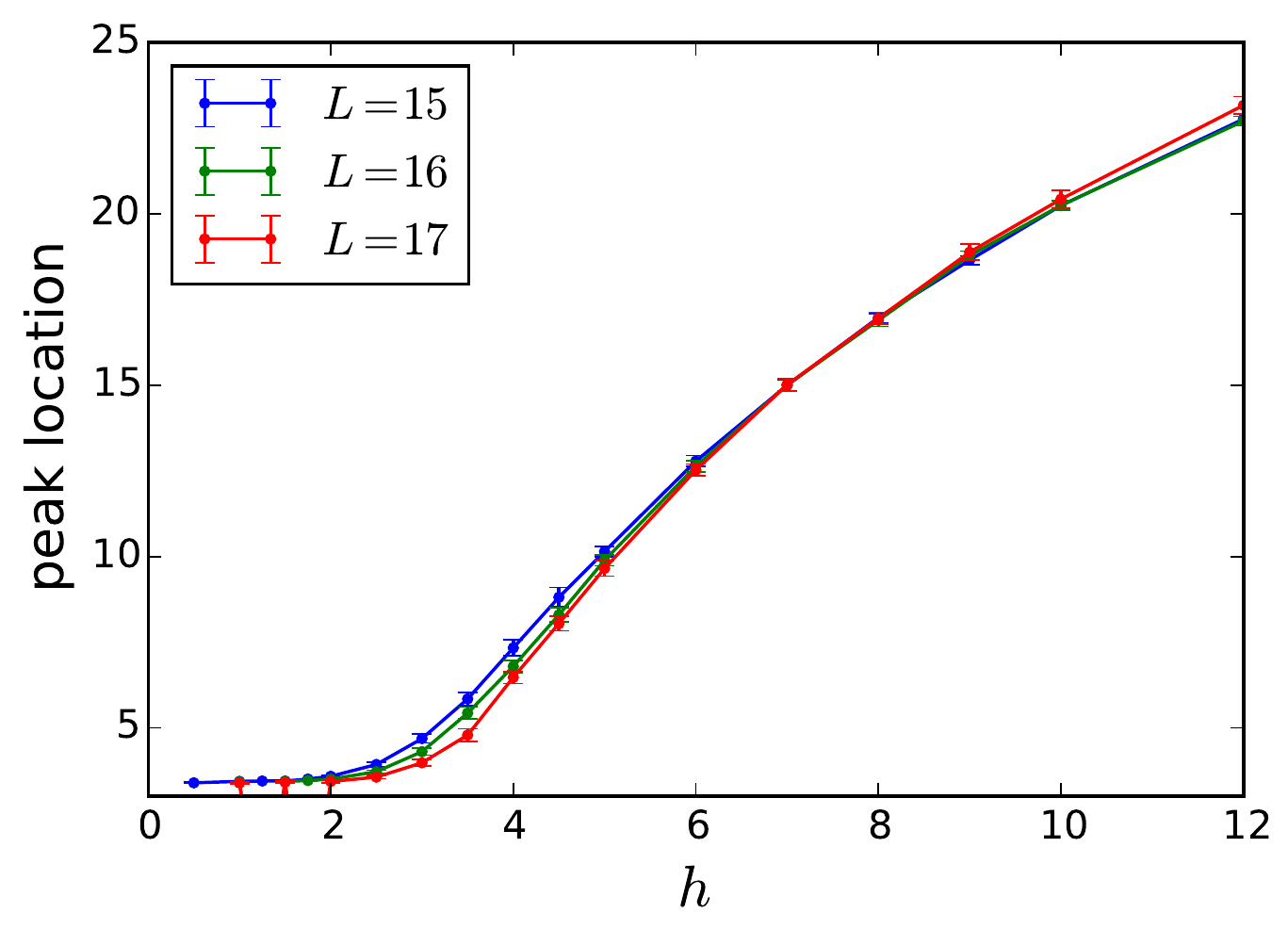}
\caption{The location of the maximum of the EDOS (excluding the peak around $\lambda=0$ which occurs in the MBL phase), as a function of $h$ for several $L$ values and $L_A=5$. In the ETH the curve is flat while the peak location is $\propto h$ in the MBL phase. 
}
\label{fig::h}
\end{figure}

We can also consider the dependence on the disorder strength $h$. Note that our toy model above suggests that in the thermal phase, the peak location should be independent of $h$ (depending only on $L_A$), whereas in the MBL phase, peak location should be $\sim 1/\xi(h)$. Now close to the transition, $\xi \sim (h-h_c)^{-\nu}$, whereas far from the transition (deep in the locator limit), $\xi \sim 1/\log(h)$. Given that exact diagonalization studies typically see $\nu \approx 0.9$ (\cite{Luitz2015}), we should expect to see the peak location be (i) independent of $h$ in the thermal regime (ii) increasing roughly linearly with $h$ close to the MBL side of the transition, and (iii) increasing much more slowly with $h$ in the deep MBL regime. In Fig.~\ref{fig::h} we show the numerically obtained dependence of peak position on $h$ for several system sizes. Note that the behavior is broadly consistent with the predictions of the toy model detailed above. 

\section{Variations between samples vs. variations between eigenstates}
\label{app::avg}

In each sample we study, every eigenstate produces its own entanglement spectrum, and therefore its own entanglement density of states. In this Appendix we compare the density of states of eigenstates coming from the same sample to those from different samples. In addition to fundamental interest, it is important to understand these difference because they inform how we perform our data analysis.

\begin{figure}[tbh]
\includegraphics[width=\linewidth]{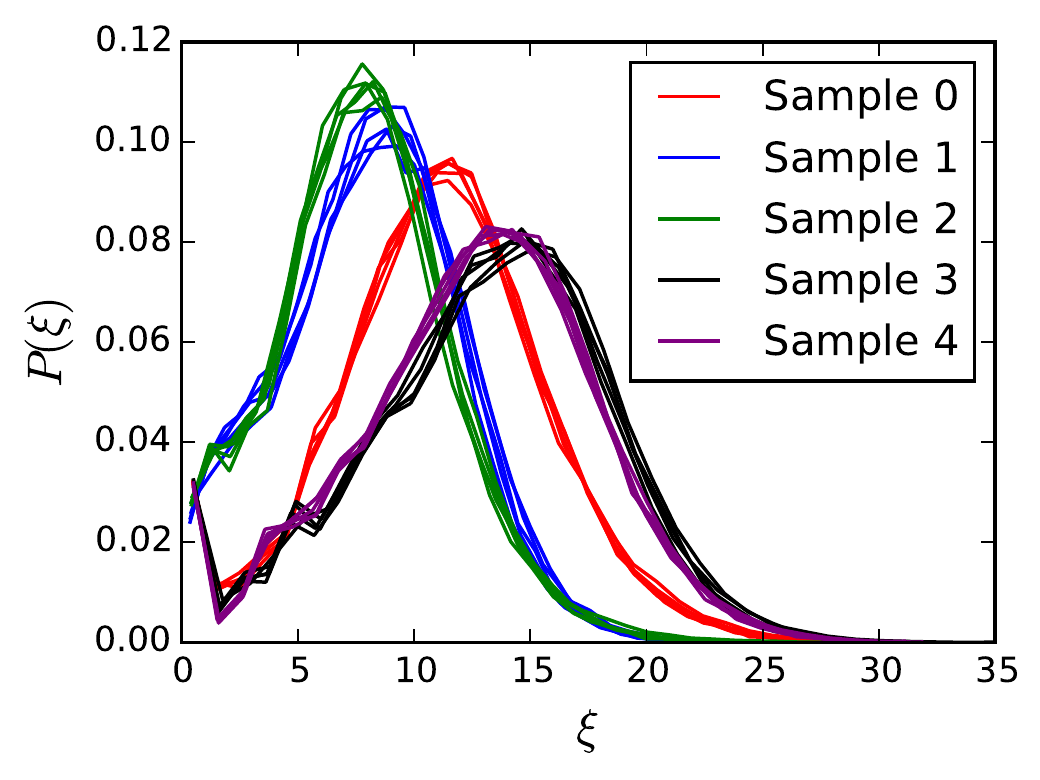}
\caption{A comparison of EDOS for different groups of eigenstates in the same sample and in different samples. The curves in the figure are obtained by averaging over $1000$ eigenstates, at $h=5$, $L_A=5$, $L=14$. Curves of the same color come from the same sample. We can see that distinct microcanonical averages of the EDOS from the same sample are much more similar than microcanonical averages from different samples, implying that sample-to-sample variation is more important than variation between microcanonical energy windows (of $1000$ states) within the same sample.}
\label{dos_seeds}
\end{figure}

Figure \ref{dos_seeds} shows such a comparison. To obtain this figure, we created a microcanonical average EDOS out of $1000$ eigenstates of a single sample. At $L=14$, where this data was taken, we focus $\approx 5000$ eigenstates (note that we restrict ourselves to the middle of the spectrum so that we do not have to worry about mobility edges), and therefore we can make five such different EDOS curves. We then repeat this procedure for multiple disorder realizations, with data coming from the same realization of disorder plotted in the same color. The plot shows that fluctuations between the EDOS obtained by microcanonical averaging over different energy windows within the same sample (different groups of 1000 eigenstates) are small compared to fluctuations between microcanonically averaged EDOS from different samples. 

\begin{figure}[tbh]
\includegraphics[width=\linewidth]{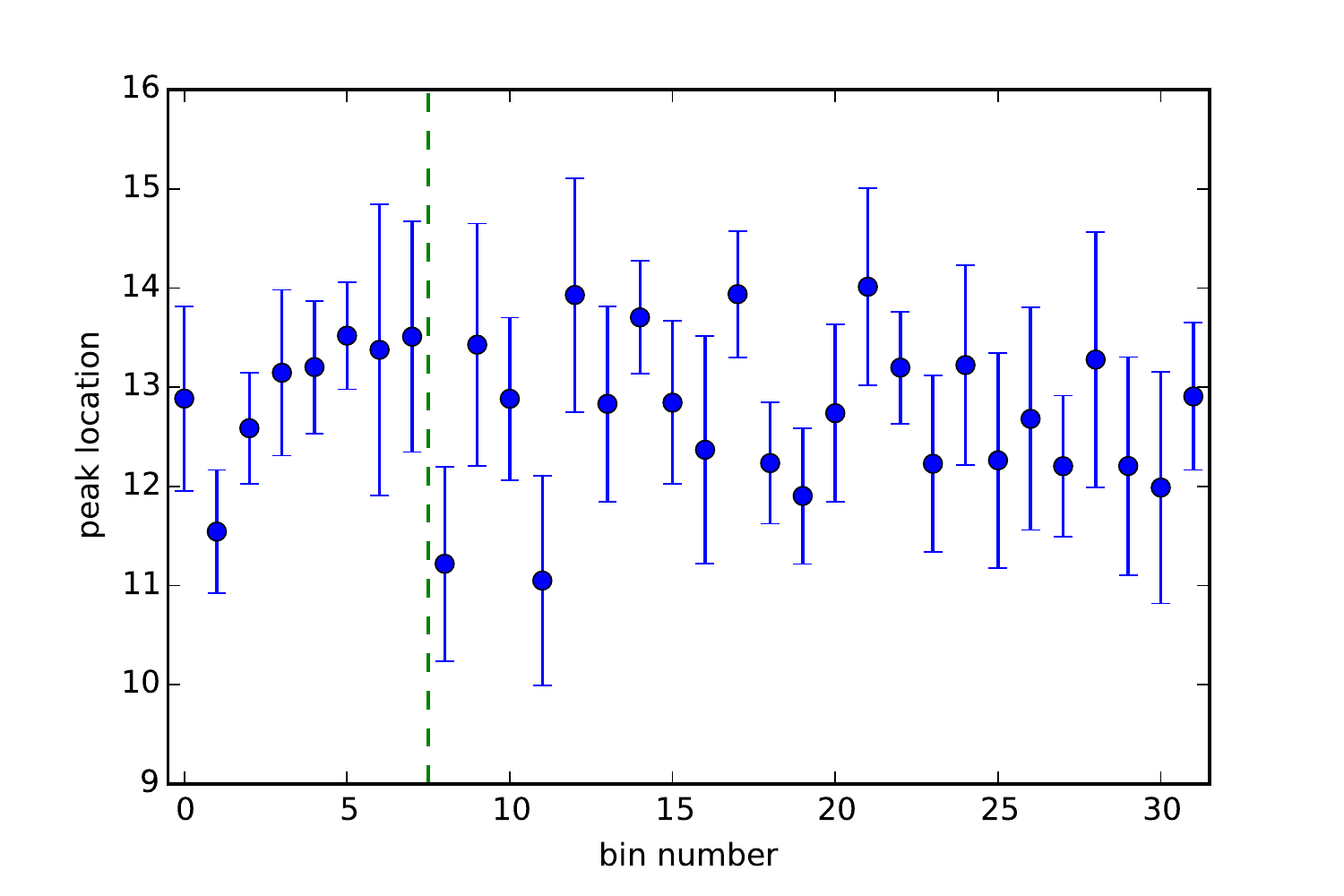}
\caption{Each data point in this figure is the result of averaging the peak locations of $16$ different samples. For the first eight points (those on the left hand side of the dashed line), all of the eigenstates of the samples were computed using full diagonalization, while in the remaining points only $1000$ eigenstates in the middle of the spectrum were computed, using the shift-invert method. The different points are consistent (within error bars), and have error bars of similar size, which leads us to conclude that our results do not depend on whether we use shift-invert or full diagonalization. Data was taken for $L=14$, $L_A=5$, $h=6$.}
\label{SIvsED}
\end{figure}

These results inform our numerical procedures in a number of ways. For large system sizes we use the shift-invert method to obtain $1000$ eigenstates in the middle of our spectrum, while for smaller system sizes we use full diagonalization. Fig.~\ref{dos_seeds} suggests that these methods should give very similar results, since the difference between EDOS obtianed from microcanonical averaging over different windows is much smaller than the fluctuations between samples. To confirm this expectation in Fig.~\ref{SIvsED} we show data for $256$ samples, where each point is the result of averaging over $16$ samples. The first four points ($64$ samples) were obtained using full diagonalization while the remainder we taken using shift-invert. If shift-invert gave different results we would expect larger error bars on the points taken with shift-invert, but we do not observe this.

\begin{figure}[tbh]
\includegraphics[width=\linewidth]{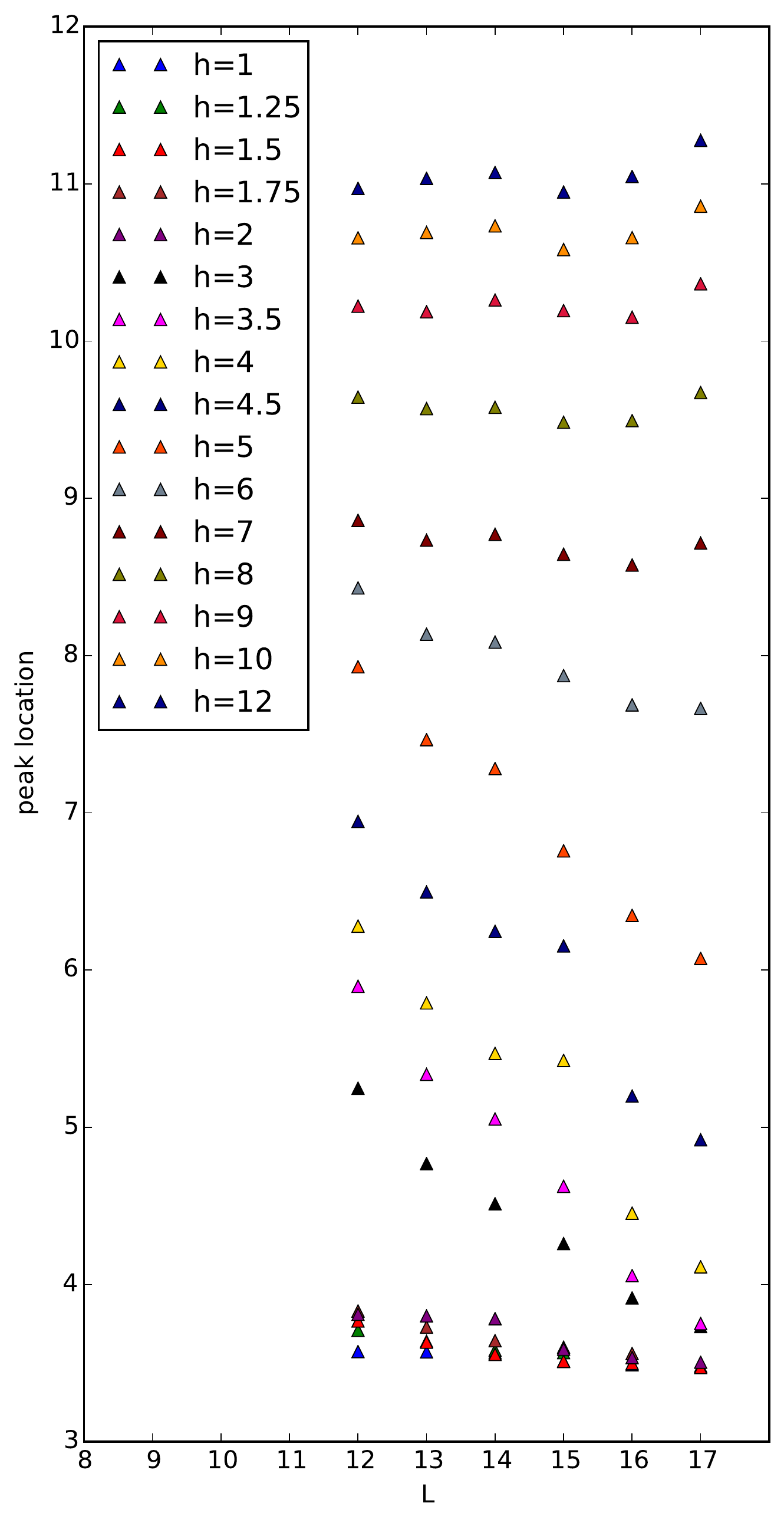}
\caption{Similar to Fig.~\ref{peaks}, but using a different method of determining the average peak location (error bars not shown). Instead of finding a peak location for each sample and averaging those, we determine one EDOS by averaging over all samples and plot its peak value. Though the results of the two methods are qualitatively the same we find that the data using the full EDOS is noisier, and moreover it is harder to estimate error bars, so we do not use this method in the main text. }
\label{Lvsh2}
\end{figure}

In the main text we compute the location of the EDOS peak by first computing an EDOS which is averaged over all eigenstates in a given sample, and then averaging those peak locations. We could imagine other ways of performing this analysis. We could compute the peak location for each eigenstate, and average those peak locations over eigenstates and samples. Fig.~\ref{dos_seeds} tells us that both of these averaging methods should give the same results, since the EDOS for all eigenstates in a given sample and the final error bar will be dominated by sample-to-sample fluctuations in any case. We could also imagine finding the peak of an EDOS which is averaged over all eigenstates and samples. The values of the peaks extracted this way will be different: in Fig.~\ref{dos_seeds} for example one can see that the results of averaging the peaks of the different samples will not necessarily be the same as the peak of the average distribution. 
In Fig.~\ref{Lvsh2} we show the results of extracting peak values from an EDOS averaged over all samples. 
To determine the values in Fig.~\ref{Lvsh2} we fit the data near the top of the curve to a Gaussian distribution, and the peak location is the maximum of this fit. Using such a method it is less clear how to assign an error bar to our data, and furthermore the data seems more noisy than that in Fig.~\ref{peaks}, and for these reasons we did not use this method in the main text. However we can clearly see that this method gives similar results to those of the main text.

\end{appendix}

\bibliography{peak}

\end{document}